\documentclass[a4paper]{jpconf}
\usepackage{graphicx}
\usepackage{xcolor}
\usepackage{amsmath}
\begin{document}
\title{Development and application of fast methods for computing momentum transfer between gas and dust in supercomputer simulation of planet formation}

\author{O P Stoyanovskaya$^1$ $^2$, V V Akimkin$^3$, E I Vorobyov$^4$ $^5$, T A Glushko$^2$ $^6$, Ya N Pavlyuchenkov$^3$, V N Snytnikov$^6$ $^2$ and N V Snytnikov$^7$}

\address{$^1$ Institute of Computational Technologies SB RAS, Novosibirsk, Russia }
\address{$^2$ Novosibirsk State University, Novosibirsk, Russia}
\address{$^3$ Institute of Astronomy RAS, Moscow, Russia}
\address{$^4$ Research Institute of Physics, Southern Federal University, Rostov-on-Don, Russia}
\address{$^5$ Department of Astrophysics, University of Vienna, Vienna, Austria}
\address{$^6$ Boreskov Institute of Catalysis SB RAS, Novosibirsk, Russia}
\address{$^7$ Institute of Computational Mathematics and Mathematical Geophysics SB RAS, Novosibirsk, Russia}

\ead{o.p.sklyar@gmail.com}

\begin{abstract}
Circumstellar discs, from which planetary systems are formed, consist of gas, dust and solids. Simulations of self-consistent dynamics of gas, dust and solids in circumstellar discs is a challenging problem. In the paper we present fast algorithms for computing the drag force (momentum transfer) between solid phase and gas. These algorithms (a) are universal and applicable to dust and solids with any sizes smaller than the mean free path of gas molecules, (b) can be used to calculate the momentum transfer between dust and gas instead of one-way effect, as it is done in many models, (c) can perform simulations, without a loss in accuracy, with the time step determined by gas-dynamic parameters rather than by drag force, and (d) are compatible with the widely used parallel algorithms for solving 3D equations of gas dynamics, hydrodynamic equations for dust, and the collisionless Boltzmann equation for large bodies. Preliminary results of supercomputer simulation of the gas-dust disc dynamics within the developed approach are reported.
\end{abstract}

\section{Introduction}
Modeling the dynamics of gas-dust circumstellar discs is a topical task of modern computational astrophysics. The hybrid model of the evolution of a gravitationally unstable circumstellar disc in which planets are formed is represented by a system of partial differential equations of general type. It includes gas dynamics equations, the Poisson equation for self-consistent gravitational field, and dust dynamics equations, which in some cases are supplemented with a subset of chemical kinetic equations and extended to take into account other processes. The modern level of computational problems and state-of-the-art approaches employed in this field is discussed in ~\cite{HaworthEtAl2016}. Our paper deals only with the numerical issues related with the simultaneous computing the gas and dust dynamics in the disc. A review of the literature on the methods for computing the gas-dust medium in circumstellar discs can be found e.g. in \cite{StoyanovskayaDust2}.

The dynamics of dust and small solids in a gaseous circumstellar disc can be described by gas dynamics equations in which the dust pressure is negligible in comparison with the gas pressure. In many disc models, the dynamics of solids is simulated with equations of the gas dynamics type separately from the dynamics of gas. In addition, dust dynamics can be described using the fully Lagrangian approach in which equations are solved for trajectories of the model or test particles, a set of which approximates the gas dynamics equations with zero pressure or a kinetic equation of the collisionless Boltzmann-Vlasov equation type. 


A general numerical issue related to time resolution is typical for all the approaches. Thus, for dust particles with the size of about 1~$\mu m$, the stopping time $t_{\rm stop}$ in the disc is about 100~s (see e.g. \cite{StoyanovskayaDust,LaibePrice2011Test}), whereas the disc dynamics should be simulated for $10^4$~years and longer. It means that the numerical solution of non-stationary equations for a multiphase gas-dust medium at times of disc dynamics using explicit integration schemes will require about $10^{10}$ time steps because each time step $\tau$ should satisfy the condition
\begin{equation}
\label{eq:timeresSPH}
\tau < t_{\rm stop}.
\end{equation}

Besides the requirements to the time step, Laibe and Price showed the necessity to use high spatial resolution when computing the dynamics of a two-phase medium by free Lagrangian methods in which gas and dust are simulated by different sets of particles. In ~\cite{LaibePrice2011Test}, Laibe and Price made test simulations of the propagation of acoustic waves in a two-phase medium using Smoothed particle hydrodynamics (SPH). Their study revealed that at a high drag factor between gas and dust and a high concentration of dust in gas, a correct computing of the perturbation amplitude can be made using a spatial step $h$ satisfying the condition
\begin{equation}
\label{eq:sparesSPH}
h<c_{\rm s} t_{\rm stop},
\end{equation}
where $c_s$ is the sound speed in gas. From (\ref{eq:sparesSPH}) it follows that for simulating the dynamics of the micron-size dust, the linear spatial size of the grid should be about $10$~km at a characteristic radius of the circumstellar disc equal to $1.5 \times 10^{10}$~km. It is clear that such spatial and time resolution is beyond the abilities of modern computers, so in practice a transition to a simplified models is widely used, for example, (1) the velocity of small grains is considered to be stationary with respect to the gas velocity or (2) the effect of the low-mass dust dynamics on the gas dynamics is neglected.

Here, we compare the earlier proposed grid \cite{StoyanovskayaDust2,VorobyovEtAl2017} and free Lagrangian methods \cite{BateDust2014,MonaghanKocharyan1995,SPHIDIC} for solving the complete problem of momentum transfer between gas and dust in a circumstellar disc. We demonstrate that both the Euler and the Lagrangian methods, in which the momentum conservation law is strictly fulfilled locally and globally, make it possible to obtain an acceptable accuracy of solution even if conditions (\ref{eq:timeresSPH}) and (\ref{eq:sparesSPH}) are violated. In addition, we present an example of computing the disc dynamics where disregard of the effect of dust dynamics on gas dynamics produces substantial differences in the dust and gas structures in the internal region of the disc.



Section \ref{sec:mathModels} presents mathematical models of gas and solid phase dynamics from small grains to large boulders. Numerical gas-dynamics methods based on the Euler and Lagrangian approaches are schematically described in section \ref{sec:numMethods}. The idea and formulas for computing the momentum and energy transfer between gas and dust in the case of Euler approach are discussed in section \ref{sec:schemes1}. Algorithms for computing the drag between gas and solid particles using the numerical models based on the SPH are reported in section \ref{sec:sph}. Modification of the parallel algorithm taking into account the momentum transfer between gas and dust is described in section \ref{sec:parallel}. Setting of the test problem and implementation details of the algorithms can be found in section \ref{sec:DustyWave}. Section \ref{sec:results} outlines the results of circumstellar disc dynamics simulation based on one of the scheme. Conclusions are listed in section \ref{sec:resume}.

\section{Mathematical models of gas and dust dynamics}
\label{sec:mathModels}
\subsection{Gas dynamics in a circumstellar disc}

The mean free path of gas molecules with respect to the pairwise collisions in a circumstellar disc is much smaller than the disc size. So the gas dynamics can be described using the Euler equations:
\begin{equation}
\label{eq:gas}
\displaystyle\frac{\partial \rho_{\rm g}}{\partial t}+\nabla (\rho_{\rm g} v)=0,\ \ \ 
 \rho_{\rm g} \left[\displaystyle\frac{\partial v}{\partial t}+(v \cdot \nabla) v \right]=-\nabla P+ \rho_{\rm g} g - f_{\rm drag},
\end{equation}
where $\rho_{\rm g}$ is the volume density of gas, $v$ is the gas motion velocity, $P$ is the gas pressure, $g$ is the gravitational acceleration, and $f_{\rm drag}$ is the drag force between dust and gas. Depending on the type of problem to be solved, this set is supplemented with the equation for specific internal energy or entropy of gas.

\subsection{A model of dust dynamics based on equations of gas dynamic type}

We consider the two-fluid polytropic model of a medium using the Euler approach in which gas and dust exchange the momentum but not the thermal energy. For submicron-size dust grains in the circumstellar disc with a large number of dust grains in a unit volume, the free path length of dust grains in their pairwise collisions is not big relative to the disc size. In this case, the volume density and the velocity of this unit volume are introduced for dust. The dynamics of the dust component in the disc will be described by Euler equations of the standard gas dynamic type of continuity and motion with zero dust pressure:
\begin{equation}
\label{eq:dust}
\displaystyle\frac{\partial \rho_{\rm d}}{\partial t}+\nabla (\rho_{\rm d} u)=0,\ \ \ 
\rho_{\rm d} \left[\displaystyle\frac{\partial u}{\partial t}+(u \cdot \nabla) u \right]=\rho_{\rm d} g+ f_{\rm drag},
\end{equation}
where $\rho_{\rm d}$ is the volume density of dust, $u$ is the velocity of dust motion, $g$ is the gravitational acceleration, and $f_{\rm drag}$ is the drag force between dust and gas.

\subsection{A model of collisionless dynamics of the dust subsystem}
When the dust grow to boulders and boulders at the times of one period around the central body do not collide with similar solids, the approach based on collisionless Boltzmann-Vlasov equation is used. In this case, instead of equations (\ref{eq:dust}) we solve the following equation: 
\begin{equation}
\label{eq:VlasovDust}
\frac{\partial{f}}{\partial{t}}+
   \mathbf{u}\frac{\partial{f}}{\partial{\mathbf{r}}}+
   (g+\textbf{f}_{\rm drag}/{m})\frac{\partial{f}}{\partial{\mathbf{u}}} = 0.
\end{equation}
Here, $f = f(t, \mathbf{r},\mathbf{u})$ is the matter distribution function (dust in the circumstellar disc), $\mathbf{u} = (u_1, u_2, u_3)$ is the velocity of dust motion, $\mathbf{r} = (r_1, r_2, r_3)$ is the coordinate of a solid grain, $m$ is the mass of a solid grain, and $\rho = \rho(t,\mathbf{r})$ denotes the density, which is calculated as 
\begin{equation}
\label{eq:VlasovRho}
\rho(t,\mathbf{r}) =
\int\limits_{\mathbf{u}}f(t,\mathbf{r}, \mathbf{u})d\mathbf{u}.
\end{equation}

\section{Numerical methods and implementation} 
\label{sec:numMethods}

\subsection{Numerical gas dynamics model based on the Euler approach}

In the numerical model FEoSaD \cite{FEoSaD} of the coevolution of a star with its circumstellar disc, a solution of the set (\ref{eq:gas})-(\ref{eq:dust}) is found by the grid-based method, which was described in detail by Stone and Norman for the case of a single-phase medium \cite{StoneNorman1992}. This method is based on operator splitting with respect to physical processes. The first stage of the splitting scheme is the computing of advective components for the mass and momentum transfer: 
\begin{equation}
\label{adv:gas}
\displaystyle\frac{\partial \rho_{\rm g}}{\partial t}+\nabla (\rho_{\rm g} v)=0,\ \ \ 
\rho_{\rm g} \left[\displaystyle\frac{\partial v}{\partial t}+(v \cdot \nabla) v \right]=0,
\end{equation}
\begin{equation}
\label{adv:dust}
\displaystyle\frac{\partial \rho_{\rm d}}{\partial t}+\nabla (\rho_{\rm d} u)=0,\ \ \ 
\rho_{\rm d} \left[\displaystyle\frac{\partial u}{\partial t}+(u \cdot \nabla) u \right]=0.
\end{equation}
This stage was implemented using the piecewise parabolic method \cite{ColellaWoodward1984}. At the second stage, the effect of drag forces and gravitation on the motion of gas and dust components of the circumstellar disc is calculated using the updated gas densities and velocities from the first stage:
\begin{equation}
\label{source:gas}
\rho_{\rm g} \frac{dv}{dt}=-\nabla P+ \rho_{\rm g} g - f_{\rm drag},
\end{equation}
\begin{equation}
\label{source:dust} 
\rho_{\rm d} \frac{du}{dt}=\rho_{\rm d} g + f_{\rm drag}.
\end{equation}
In this case, the terms in square brackets in the equations (\ref{eq:gas})-(\ref{eq:dust}) are presented as the total time derivatives.

\subsection{Numerical model with particles based on the Lagrangian approach}

In the numerical code Sombrero \cite{1MNRAS,ASCOM2017}, gas dynamics equations (\ref{eq:gas}) are solved by the free Lagrangian SPH method. According to this method, a continuous medium is represented by a set of densely packed particles, which are the carriers of main characteristics of the medium -- mass, momentum, and energy. 

When a two-phase system is described using the gas dynamic approximation, equations of the model particles motion also have the form (\ref{source:gas})-(\ref{source:dust}). If the collisionless approximation is used to describe the solid-phase, the Boltzmann-Vlasov equation is solved by the particle-in-cell method. In our case, we approximate the distribution function by quite a large number of particles, which reaches several thousand particles in a cell for modern supercomputers. The assigned masses and velocities of the particles together with one of the smoothing algorithms in the grid nodes are used to calculate the total density, velocity and other quantities, which are the momenta of the Boltzmann-Vlasov equation. Particles move according to the forces that are calculated by interpolation of the found forces in the cell nodes to the point where the particle resides.

\section{Computing of the rapid momentum transfer between gas and solids}
\label{sec:computing}

Let us consider a medium in which solids interact with gas in the Epstein regime or in the regime of free molecular flow, which means that 
\begin{equation}
\label{eq:fdrag}
f_{\rm drag}=\rho_{\rm d} \displaystyle \frac{v-u}{t_{\rm stop}}.
\end{equation}

Fig. 2 in \cite{StoyanovskayaDust} shows a maximum size of the dust particles in the disc that can be described using this approximation. 

\subsection{The idea of the method and formulas for the Euler approach}
\label{sec:schemes1}

Assume that $\varepsilon=\displaystyle\frac{\rho_{\rm d}}{\rho_{\rm g}}$ is the mass fraction of dust with respect to gas, 
\begin{equation}
a_{\rm g}=-\displaystyle\frac{\nabla P}{\rho_{\rm g}}+g, \ \ a_{\rm d}=g
\end{equation}
are the accelerations, except drag acceleration acting on the gas and dust, then the equations (\ref{source:gas})-(\ref{source:dust}) take the form:
\begin{equation}
\label{eq:system}
        \displaystyle \frac{dv}{dt} = a_{\rm g}-\varepsilon\displaystyle\frac{v-u}{t_{\rm stop}}, \quad
        \displaystyle 
        \frac{du}{dt} = a_{\rm d} + \displaystyle\frac{v-u}{t_{\rm stop}}. 
\end{equation}

Suppose we know the velocities of the selected gas and dust volume at a certain time point: 
\begin{equation}
\label{eq:systemInit}
v|_{t=t_0}=v_n, u|_{t=t_0}=u_n.
\end{equation}

Let us find the velocities $v^{n+1}$ and $u^{n+1}$ of the same gas and dust volume after a time step $\tau$. 

\subsubsection{The SIOS semi-implicit scheme with operator splitting.}

The SIOS numerical scheme was constructed by analogy with the semi-implicit scheme from \cite{ChaNayakshin2011}

\begin{equation}
\label{eq:InitFinite}
\displaystyle\frac{v^{n+1}-v^{n}}{\tau}=a^n_{\rm g}-\varepsilon\displaystyle\frac{v^{n+1}-u^n}{t_{\rm stop}}, \quad 
\displaystyle\frac{u^{n+1}-u^n}{\tau}=a^n_{\rm d}+\frac{v^{n+1}-u^{n+1}}{t_{\rm stop}}.\\
\end{equation}

At $a_{\rm g}=0$, $a_{\rm d}=0$ and $\varepsilon=const$ the momentum conservation law in each selected volume for (\ref{eq:system}) has the form  
\begin{equation}
\label{eq:dragTransfer}
\displaystyle\frac{\partial}{\partial t}(v+\varepsilon u)=0.
\end{equation}

Instead of a discrete analogue of the momentum conservation law $v^{n+1}+\varepsilon u^{n+1}=v^n+\varepsilon u^n$, the following relation takes place:
\begin{equation}
v^{n+1}+\varepsilon u^{n+1}=v^n+\varepsilon u^n+\displaystyle\frac{\tau}{t_{\rm stop}}\varepsilon (u^n-u^{n+1}).
\end{equation}

\subsubsection{The EMSI semi-implicit scheme ensuring the conservation of momentum.}

To construct a scheme in which a discrete analogue of the momentum conservation law is fulfilled in each cell, let us turn to the equivalent set of equations for relative $x=v-u$ and barycentric $y=v+\varepsilon u$ velocities of the gas-dust medium:
\begin{equation}
\label{eq:Newsystem}
     \displaystyle 
     \frac{dx}{dt} = (a_{\rm g}-a_{\rm d})-\displaystyle\frac{\varepsilon+1}{t_{\rm stop}}x, \quad
        \displaystyle 
        \frac{dy}{dt} = a_{\rm g}+\varepsilon a_{\rm d}. 
\end{equation}

Approximating the first equation by the implicit scheme of the first order approximation with respect to time and recording an exact solution of the second one, we will obtain the formulas:

\begin{equation}
\label{eq:Finite}
        x^{n+1} = \displaystyle \frac{x^n+\tau(a_{\rm g}-a_{\rm d})}{1+(\varepsilon+1)\displaystyle\frac{\tau}{t_{\rm stop}}}, \quad
        y^{n+1} = y^n+\tau (a_{\rm g}+\varepsilon a_{\rm d}). 
\end{equation}

\begin{equation}
\label{eq:FiniteVU}
        v^{n+1} = \displaystyle \frac{\varepsilon x^{n+1}+y^{n+1}}{\varepsilon+1}, \quad 
        u^{n+1} = \displaystyle \frac{y^{n+1}-x^{n+1}}{\varepsilon+1}. 
\end{equation}
These formulas can be explicitly applied to the Euler approach, in which spatial grids for gas and dust coincide.

\subsection{A method for computing the drag in two-fluid Smoothed particle hydrodynamics}
\label{sec:sph}

Let us rewrite the set (\ref{source:gas})-(\ref{source:dust}) assuming $K=\displaystyle\frac{\rho_{\rm d}}{t_{\rm stop}}$:
\begin{equation}
\label{eq:systemK}
\left\{
 \begin{array}{lcl}
        \displaystyle 
        \frac{dv}{dt} = -\displaystyle \frac{\nabla P}{\rho_{\rm g}} + g -\displaystyle\frac{K}{\rho_{\rm g}} (v - u), \\
        \displaystyle 
        \frac{du}{dt} = g + \displaystyle\frac{K}{\rho_{\rm d}} (v - u). 
    \end{array}
\right.
\end{equation}

Further we will give the schemes for solving the equations of gas and dust motion (\ref{eq:systemK}) in the notation that is standard for SPH. We will consider only the schemes in  which gas and dust are simulated by different sets of particles, i.e. by the two-fluid approach for Smoothed particle hydrodynamics (TFSPH). In comparison with the grid methods examined in the previous sections, the distinctive feature of TFSPH is that the drag force depends on characteristics of the gas and dust medium, whereas characteristics of either gas or dust medium are known for each model particle. Let $n$ be the number of the time step. Following the notations introduced in \cite{MonaghanKocharyan1995}, we will use $a,b$ as the indices for gas particles, and $i,j$ as the indices for dust particles.

\subsubsection{The MKD Monaghan - Kocharyan explicit scheme.}
A method for computing the drag force, which was proposed in \cite{MonaghanKocharyan1995}  (hereinafter referred to as MKD (Monaghan-Kocharyan Drag)), is classical for Smoothed particle hydrodynamics. This method is based on the computing of the relative velocity between each pair of gas-dust particles and is employed in astrophysical and engineering applications of two-phase medium mechanics \cite{Maddison2004TFSPH,FranceDustCode,Gonzalez2017,TFSPH} and others.

We implemented this scheme so that the term accounting for drag includes the velocities from the previous time step: 
\begin{equation}
\displaystyle\frac{dv^n_a}{dt}= - m_{\rm g} \sum_b \left(\frac{P_b}{(\rho^n_{b,\rm{g}})^2} + \frac{P_a}{(\rho^n_{a,\rm g})^2} \right) \bigtriangledown_a W^{n}_{ab} 
- \sigma m_{\rm d} \sum_j \frac{K_{aj}}{\rho^n_{a, \rm g} \rho^n_{j,\rm d}} \frac{(v_a^{n} - u_j^{n},r^n_{ja})}{(r^n_{ja})^2+\eta^2}r^n_{ja}W^{n}_{ja}
+g_a,
\label{eq:MonKoch1995v}
\end{equation}
\begin{equation}
\displaystyle\frac{du^n_j}{dt}= \sigma m_{\rm g} \sum_a \frac{K_{aj}}{\rho^n_{a, \rm g} \rho^n_{j,\rm d}} \frac{(v_a^{n} - u_j^{n},r^n_{ja})}{(r^n_{ja})^2+\eta^2}r^n_{ja}W^{n}_{ja}+g_j,
\label{eq:MonKoch1995u}
\end{equation}
\begin{equation}
K_{aj}=\displaystyle\frac{\rho^n_{j,\rm d} \rho^n_{a, \rm g} c^n_{a,\rm s}}{s_j^n \rho^n_{j,\rm s}},
\end{equation}
where $m_{\rm g}$ and $m_{\rm d}$ are the masses of gas and dust particles, respectively, $r^n_{ja} = r^n_j - r^n_a$, $\eta$ is a clipping constant, $\eta^2 = 0.001 h^2$, $s_j$ is the radius of a spherical dust particle with index $j$, $\sigma$ is the constant determined by dimensionality of the problem (for one-dimensional problems, $\sigma=1$), and $W^{n}_{ab}=W(h,r^n_{ab})$ is the smoothing kernel.

The MKD-SPH scheme (\ref{eq:MonKoch1995v})-(\ref{eq:MonKoch1995u}) of the first order approximation with respect to time satisfies the momentum conservation law (\ref{eq:dragTransfer}) in the entire computational domain, which means that the momentum lost by gas due to dust drag completely coincides with the momentum acquired by dust due to gas drag.

\subsubsection{The explicit ESPH scheme with interpolation.}

The second method for computing the drag consists in the computing of gas characteristics at the points where dust particles are located (and vice versa) using the SPH interpolation formulas:
\begin{equation}
\label{eq:VUvolaverage}
v^n_j=\displaystyle m_{\rm g} \sum_a \frac{v^n_a}{\rho^n_{a, \rm g}} W^n_{aj}, \quad u^n_a = \displaystyle m_{\rm d} \sum_j \frac{u^n_j}{\rho^n_{j, \rm d}} W^n_{ja}, 
\end{equation}
where $u_a$ is the dust velocity at a spatial point where the gas particle with $a$ index  is located, and $v_j$ is the gas velocity at a spatial point where the dust particle with $j$ index is located.

As a result, all features of the gas-dust medium become known for each model particle. This method and its modifications are applied in \cite{BateDust2014,ClarkeDust2015,RiceEtAl2004}. The explicit ESPH scheme with interpolation is represented by the formulas (the quantities calculated using interpolation formulas are marked in blue, while the quantities derived from those calculated by interpolation formulas are marked in red):
\begin{equation}
\label{eq:VESPH}
\displaystyle\frac{dv^n_a}{dt}= - \sum_b m_b\left(\frac{P_b}{(\rho^n_{b,\rm{g}})^2} + \frac{P_a}{(\rho^n_{a,\rm g})^2} \right) \bigtriangledown_a W^{n}_{ab} - 
\frac{\textcolor{red}{K_a^n}}{\rho^n_{a,\rm g}} (v_a^{n} - \textcolor{blue}{u_{a}^{n}})+g_a,
\end{equation}
\begin{equation}
\label{eq:UESPH}
\displaystyle\frac{du^n_j}{dt}= \frac{\textcolor{red}{K_j^n}}{\rho^n_{j,\rm d}} (\textcolor{blue}{v^{n}_j} - u^{n}_j)+g_j,
\end{equation}
\begin{equation}
\label{eq:KintSPH}
\textcolor{red}{K^n_a}=\displaystyle\frac{\rho^n_{a, \rm g} c^n_{a,\rm s}}{\textcolor{blue}{s_a^n \rho^n_{a,\rm s}}}, \quad 
\textcolor{red}{K^n_j}=\displaystyle\frac{\textcolor{blue}{\rho^n_{j, \rm g} c^n_{j,\rm s}}}{s_j^n \rho^n_{j,\rm s}}.
\end{equation}

\subsubsection{The first order in time semi-implicit ISPH scheme with interpolation.}

The idea of interpolation can be used to construct also a semi-implicit scheme that would not require the fulfillment of condition (\ref{eq:timeresSPH}) for obtaining stable solutions. In particular, the following ISPH scheme with the first order approximation with respect to time is fast:
\begin{equation}
\label{eq:VISPH}
\displaystyle\frac{v^{n+1}_a-v^n_a}{\tau}= - \sum_b m_b\left(\frac{P_b}{(\rho^n_{b,\rm{g}})^2} + \frac{P_a}{(\rho^n_{a,\rm g})^2} \right) \bigtriangledown_a W^{n}_{ab} - 
\frac{\textcolor{red}{K_a^n}}{\rho^n_{a,\rm g}} (v_a^{n+1} - {u_{a}^{n+1}})+g_a,
\end{equation}
\begin{equation}
\displaystyle\frac{u^{n+1}_a-\textcolor{blue}{u^n_a}}{\tau}= \frac{\textcolor{red}{K_a^n}}{\textcolor{blue}{\rho^n_{a,\rm d}}} (v^{n+1}_a - u^{n+1}_a)+g_a.
\end{equation}
\begin{equation}
\label{eq:UISPH}
\displaystyle\frac{v^{n+1}_j-\textcolor{blue}{v^n_j}}{\tau}= - \sum_i m_i\left(\textcolor{blue}{\frac{P_i}{(\rho^n_{i,\rm{g}})^2} + \frac{P_j}{(\rho^n_{j,\rm g})^2}} \right) \bigtriangledown_j W^{n}_{ij} - 
\frac{\textcolor{red}{K_j^n}}{\textcolor{blue}{\rho^n_{j,\rm g}}} (v_j^{n+1} - {u_{j}^{n+1}})+g_j,
\end{equation}
\begin{equation}
\displaystyle\frac{u^{n+1}_j-u^n_j}{\tau}= \frac{\textcolor{red}{K_j^n}}{\rho^n_{a,\rm d}} ({v^{n+1}_j} - u^{n+1}_j)+g_j.
\end{equation}

It is seen that, in distinction to the explicit ESPH scheme, implementation of the proposed scheme makes it necessary to solve a set of two discrete analogues of equations (\ref{eq:Newsystem})-(\ref{eq:FiniteVU}) for each gas and dust model particle instead of one equation of motion.

\subsubsection{A new SPH-IDIC scheme -- the implicit ``drag in cell''.}

In addition, computing of the drag force can be based on the idea of the Particle-in-Cell method for simulation of gas-dust flows \cite{Andrews1996}. The fast semi-implicit SPH-IDIC approach based on this idea was suggested and tested in our earlier paper \cite{SPHIDIC}. A detailed description of this approach is presented below. 

At each time step, we will split the entire computational domain into closely spaced volumes so that the merging of these volumes will coincide with the entire region. Suppose a separate volume contains $N$ gas particles of a equal mass $m_{\rm g}$ and $L$ dust particles of a equal mass $m_{\rm d}$, with $N>0$, $L>0$. Introduce the volume-averaged values of $t^*_{\rm stop}$ and $\rho^*_{\rm d}$ (anywise) and assume that 
\begin{equation}
\label{eq:aveEpsilon}
\varepsilon^*=\displaystyle\frac{m_{\rm d} L}{m_{\rm g} N}, 
\end{equation}
thus determining 
\begin{equation}
K^*=\frac{\rho^*_{\rm d}}{t^*_{\rm stop}}, \ \ \rho^*_{\rm g}=\frac{\rho^*_{\rm d}}{\varepsilon^*}. 
\end{equation}

Let us assume that in computing the drag force that acts from gas on dust, the gas velocity is constant over the entire volume and equal to $v_*$, whereas dust particles have different velocities (and vice versa). 
In addition, we will calculate the drag coefficient and density using values of the quantities from the previous time step, and relative velocity -- from the next step. The resulting scheme will have the form 
\begin{equation}
\displaystyle\frac{dv^n_a}{dt}= - \sum_b m_b\left(\frac{P_b}{(\rho^n_{b,\rm{g}})^2} + \frac{P_a}{(\rho^n_{a,\rm g})^2} \right) \bigtriangledown_a W^{n}_{ab} - \frac{K^*}{\rho^{n,*}_{\rm g}} (v_a^{n+1} - u_{*}^{n+1})+g_a,
\label{eq:DragInCellv}
\end{equation}
\begin{equation}
\displaystyle\frac{du^n_j}{dt}= \frac{K^*}{\rho^{n,*}_{\rm d}} (v^{n+1}_* - u^{n+1}_j)+g_j,
\label{eq:DragInCellu}
\end{equation}
\begin{equation}
\label{eq:VUvolaverage}
v_*=\displaystyle\frac{\sum_{a=1}^N v_a}{N}, \quad u_*=\displaystyle\frac{\sum_{j=1}^L u_j}{L}.
\end{equation}

If the time derivative in (\ref{eq:DragInCellv})-(\ref{eq:DragInCellu}) is approximated to the first order, there exists a fast method to calculate $u^{n+1}$ $v^{n+1}$, which is similar to that described in section (\ref{sec:schemes1}) and will be reported in section \ref{realizeIDIC}. In \cite{SPHIDIC}, it was shown that the semi-implicit scheme (\ref{eq:DragInCellv})-(\ref{eq:VUvolaverage}) with the first order approximation with respect to time satisfies the momentum conservation law (\ref{eq:dragTransfer}) for each cell, i.e. the momentum lost by gas due to dust drag completely coincides with the momentum acquired by dust due to gas drag.

\section{A parallel algorithm for collisionless dynamics of solids and gas}
\label{sec:parallel}

The parallel algorithm developed for this numerical model is based on the approaches proposed earlier in refs.~\cite{ncc2015,ugatu,pavt17} for supercomputers with distributed memory and takes into account the presence of two types of particles (collisionless particles and SPH ones).

The essence of the algorithm is that the rectangular computational domain is decomposed into $N$ similar regions. A certain group of processors is assigned to each subregion: the ``main'' processor in each group is used to compute the Poisson equation, whereas the other processors are used for processing of SPH particles and collisionless dust particles. 


Modifications required for tracking the trajectories of particles in the presence of drag force are based on the following main ideas:
\begin{itemize}
\item The algorithm for computing the drag force implies that the closely located SPH particles and collisionless particles should reside on the same processor in order to avoid data transfer. This requires the application of specialized algorithms of particle sorting relative to a cell and constructing K-d trees in each cell if the number of particles in one cell is quite large.  
\item The computational costs on integrating trajectories of the particles of both types and calculate the drag force between them strongly exceeds (100-fold and more) the costs on computing of the gravitational potential.
\item For advanced supercomputers, the transfer of a moderate amount of data between adjacent processors (coordinates and parameters of 10-100 thousand particles) takes a much shorter time than the integration of trajectories of the particles. 
\item According to the Courant condition, particles of both types can move only between adjacent cells. Thus, the amount of data required for the transfer between processors constitutes no more than 1\% of the total number of particles. This implies the introduction of the so-called boundary regions with overlap (``ghost zones''), where information on the particles of both types is duplicated for adjacent processors. 
\end{itemize}

Preliminary experiments were carried out with the numerical model of a gas-dust medium taking into account the drag force between dust and gas. For $100$ million SPH particles and $1000$ million collisionless particles on a $16384 \times 16384$ grid using 128 kernels, the computing of one time step took about 40 seconds.

\section{Test. Sound waves in a two-phase isothermal medium DustyWave}
\label{sec:DustyWave}
\subsection{Statement of the problem. A reference solution}
Assume that gravitational acceleration of gas and solids, $P=c_s^2 \rho_{\rm g}$, is absent in the system (\ref{eq:gas})-(\ref{eq:dust}), while the sound velocity is taken to be constant. Let us consider a solution on the interval $x \in [0,1]$ with non-negative sound speed, setting the periodic values of functions with respect to $x$ for solutions on the left boundary: 
\begin{equation}
\label{eq:periodic}
\rho_{\rm g}|_{x=0}=\rho_{\rm g}|_{x=1}, \ \ \rho_{\rm d}|_{x=0}=\rho_{\rm d}|_{x=1}, \ \ v|_{x=0}=v|_{x=1}, \ \ u|_{x=0}=u|_{x=1},
\end{equation}
and initial data as a perturbation of stationary density and velocity:
\begin{equation}
\label{eq:DustyWave_init1}
\rho_{\rm g}|_{t=0}=\tilde{\rho_{\rm g}}+\delta \sin(kx), \ \rho_{\rm d}|_{t=0}=\tilde{\rho_{\rm d}}+\delta \sin(kx), 
\end{equation}
\begin{equation}
\label{eq:DustyWave_init2}
v|_{t=0}=\delta \sin(kx), \ u|_{t=0}=\delta \sin(kx).
\end{equation}
Here, $k$ is the wavenumber specifying an integer number of sine waves of density and velocity on the interval $[0,1]$, and $\delta$ is the perturbation amplitude. In the vicinity of a stationary point, the linearized system (\ref{eq:gas})-(\ref{eq:dust}) will take the form:
\begin{equation}
\label{eq:LinDustyWaveCont}
\frac{\partial \delta \rho_{\rm g}}{\partial t}+\tilde{\rho_{\rm g}} \frac{\partial{v}}{\partial x} = 0, \ \  
\frac{\partial \delta \rho_{\rm d}}{\partial t}+\tilde{\rho_{\rm d}} \frac{\partial{ u}}{\partial x} = 0,\ \ 
\end{equation}
\begin{equation}
\label{eq:LinDustyWaveMotionGas}
\tilde {\rho_{\rm g}} \frac{\partial v}{\partial t} = - c_s^2 \frac{\partial \delta \rho_{\rm g}}{\partial x} - K(v-u),
\end{equation}
\begin{equation}
\label{eq:LinDustyWaveMotionDust}
\tilde {\rho_{\rm d}} \frac{\partial u}{\partial t} =  K(v-u).
\end{equation}

The analytical solution of the linearized system (\ref{eq:LinDustyWaveCont})-(\ref{eq:LinDustyWaveMotionDust}) can be found in \cite{LaibePrice2011}. The authors of \cite{LaibePrice2011} also provided free access to the code for generation of this solution, and we used it in our work.
For simplicity, we will further refer to the exact solution of the linearized set (\ref{eq:LinDustyWaveCont})-(\ref{eq:LinDustyWaveMotionDust}) as the analytical solution of the (\ref{eq:gas})-(\ref{eq:dust}) system. 

In the reported runs, the initial distributions of gas and dust velocity are specified as $u_0 = v_0 = \displaystyle \delta \sin(2\pi x)$, density of gas as $\rho_{\rm g,0} = \displaystyle \delta \sin(2\pi x) + 1$, and density of dust as $\rho_{\rm d,0} = \displaystyle \delta \sin(2\pi x) + \varepsilon$ with parameters $\delta = 10^{-4}$, $c_s = 1$, $\varepsilon = 1$.

\subsection{Details of implementing the Smoothed particle hydrodynamics}

Continuity equations for dust and gas are approximated by the standard SPH interpolation:
\begin{equation}
\label{eq:contin_gas}
\rho^{n}_{a, \rm g} = m_{\rm g} \sum_b W^{n}_{ab},
\end{equation}
\begin{equation}
\label{eq:contin_dust}
\rho^{n}_{i, \rm d} = m_{\rm j} \sum_i W^{n}_{ij}.
\end{equation}

A cubic spline is used as the kernel:
\begin{equation}
\label{eq:kernel}
W^{n}_{ab} = W(|r^{n}_a - r^{n}_b|, h) = W^{n}(q) = \frac{2}{3 h}
 \begin{cases}
   1 - \frac{3}{2} q^2 + \frac{3}{4} q^3, &\text{if $ \displaystyle 0 \le q \le 1$,}\\
   \displaystyle \frac{1}{4}(2 - q)^3, &\text{if $\displaystyle 1 \le q \le 2$,}\\
   0, &\text{otherwise;}
 \end{cases}
\end{equation}
where $q = \displaystyle \frac{|r^{n}_a - r^{n}_b|}{h}$.

At the spatial points where gas particles are located, the pressure is calculated as $P_a = c^2_s \rho_{a, \rm g}$, and artificial viscosity is not introduced.

Equations of motion are integrated with respect to time with the first order using the ``backward difference'', which means that $\displaystyle \frac{df^n}{dt} = \frac{f^{n+1} - f^n}{\tau}$ for an arbitrary function $f$. Therewith, the time step $\tau$ is determined from the Courant condition:
\begin{equation}
\label{eq:courant}
\tau < \displaystyle \frac{h \cdot CFL}{max(c_s, u, v)},
\end{equation}
where $CFL$ is the Courant parameter.

To obtain on a segment $[0,l]$ the initial density perturbed with respect to a constant value $\tilde{\rho}$ by $\delta \sin(2\pi x)$, i.e., $\rho_0(x)=\tilde{\rho}+\delta \sin(2\pi x)$, we used a recurrent procedure for arranging the model particles. The first particle was placed at the origin of coordinates $x_1=0$, and a coordinate of the next particle was found from the relation   

$$\displaystyle \int\limits_{x_i}^{x_i + \Delta x_i} \rho_0 dx = \frac{\tilde{\rho} l}{N_{\rm ph}},$$
where $N_{\rm ph}$ is the number of model gas or dust particles. After finding all $x_i$, each particle was shifted to the right by the $\Delta x_i$ value.

At the initial time moment, particles are arranged on a segment [0:1], and coordinates of gas and dust particles coincide. To ensure the periodic boundary conditions (\ref{eq:periodic}), at each time step the particles from the segment under consideration were duplicated to the right and to the left by a length equal to the segment length.







\subsubsection{Implementation of the SPH-IDIC scheme with the first order with respect to time.}
\label{realizeIDIC}

To calculate the drag force by the SPH-IDIC scheme, the entire computing region was decomposed into segments of equal length $h_{\rm cell} = h$. In each cell, $\rho^*_{\rm d}$ was calculated as $\rho^*_{\rm d} = \displaystyle \frac{\sum_{j=1}^L \rho_{\rm d}}{L}$, $t^*_{\rm stop} = \displaystyle \frac{\rho^*_{\rm d}}{K}$.

Let us denote $\Psi^{n}_a  = - \displaystyle \sum_b m_b\left(\frac{P^{n}_b}{(\rho^n_{b,\rm{g}})^2} + \frac{P^{n}_a}{(\rho^n_{a,\rm g})^2} + \Pi_{ab} \right) \bigtriangledown_a W_{ab}$.

Summarize the equation (\ref{eq:DragInCellv}) over $a$ and divide the obtained value by the number of gas particles $N$ in a cell. This will give the equation over $v_*$:
\begin{equation}
\label{eq:v_asterisk}
\frac{v^{n+1}_*-v^{n}_*}{\tau} = \displaystyle \Psi^n_* - \varepsilon^n_* \frac{v^{n+1}_* - u^{n+1}_*}{t^{n, *}_{stop}},
\end{equation}
where $\Psi^n_* = \displaystyle \frac{1}{N} \sum^N_{i=1} \Psi^n_a$. 

Similarly, for $u_*$:
\begin{equation}
\label{eq:u_asterisk}
\frac{u^{n+1}_*-u^{n}_*}{\tau} = \displaystyle \frac{v^{n+1}_* - u^{n+1}_*}{t^{n, *}_{\rm stop}}.
\end{equation} 

Then the set (\ref{eq:v_asterisk})-(\ref{eq:u_asterisk}) is reduced to the set (\ref{eq:Newsystem}) by the substitution $x = v - u, \ \ y = v + \varepsilon u$ and is solved using the formulas (\ref{eq:Finite})-(\ref{eq:FiniteVU}).

\section{Results of computing}
\label{sec:results}

\subsection{Investigation of the properties of the drag computing algorithms}

\begin{figure}[h]
\center
  \includegraphics[scale=0.17]{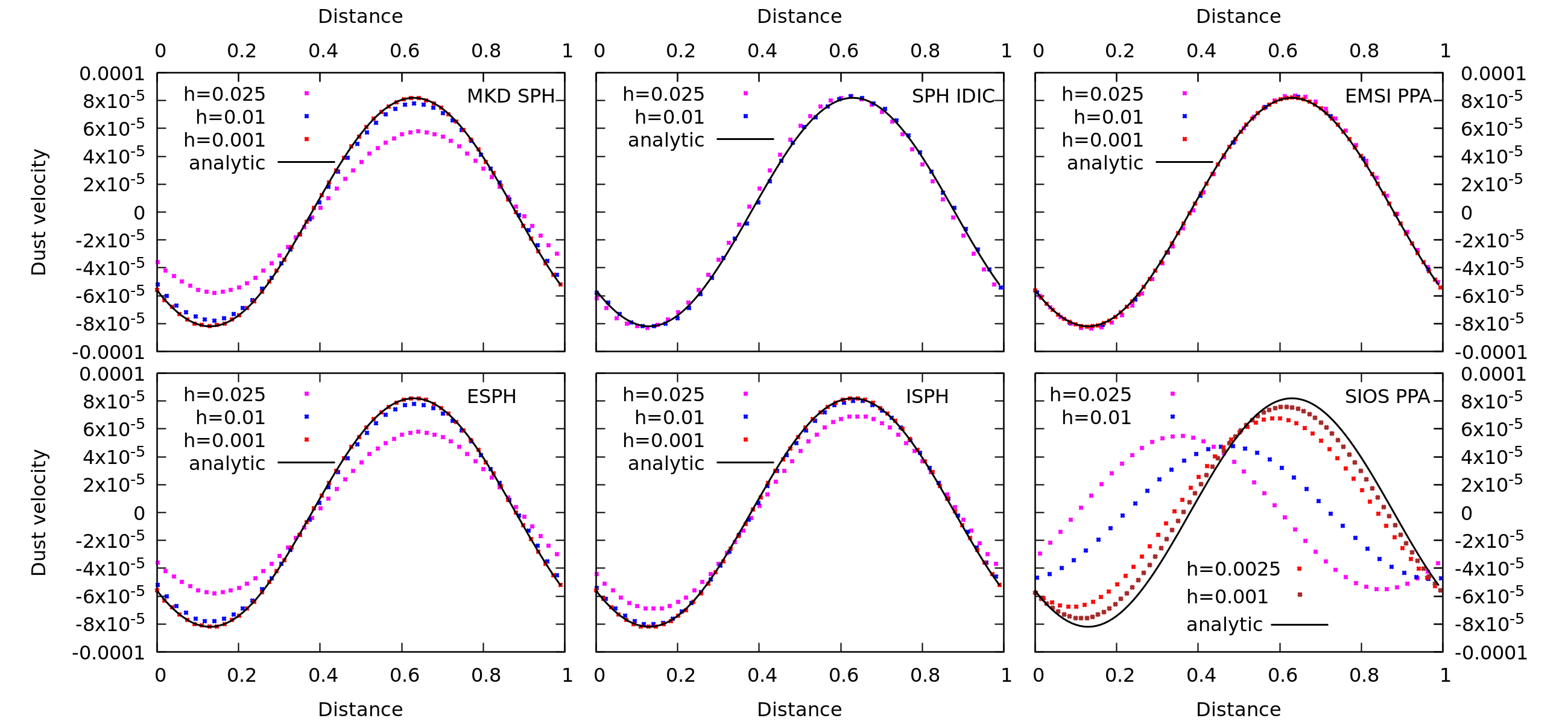}
\caption{Solution of the DustyWave problem at the time instant $t=0.5$ for a gas-dust medium with a high dust content $\varepsilon=1$ and a short relaxation time of the dust velocity with respect to gas $t_{\rm stop}=0.002$, i.e. $t_{\rm stop} c_{\rm s} / l \ll 1$, where $l$ is the length of the computing region. Solid black line corresponds to the analytical solution, and individual dots are the numerical solutions. The left panels show the explicit methods based on SPH; at $h = 0.025$ and $h=0.01$, the time step $\tau = 0.001<t_{\rm stop}$ and the number of particles $N_{total} = 2 \times 600$ are used; at $h=0.001$, the step is $\tau = 0.0001$ and $N_{total} = 2 \times 6000$. The middle panels display semi-implicit methods based on SPH with $CFL=0.1$; the number of particles is the same as for the left panels. The right panels show the grid methods with $CFL=0.5$.}
\label{fiq:schemes}
\end{figure}

Fig.~\ref{fiq:schemes} illustrates the operation of the schemes described in section \ref{sec:computing}, which are applied to the DustyWave test problem at the time moment $t = 0.5$ at a high drag coefficient $K=500$, a high concentration of dust in gas $\varepsilon = 1$ ($t_{stop} = 0.002$), and a small amplitude of the initial perturbation $\delta = 10^{-4}$. The MKD, ESPH, ISPH, SIOS PPA, and EMSI PPA schemes were implemented with a constant drag coefficient $K=const$. The figure displays the dust velocity obtained with different smoothing lengths $h$ (SPH-based methods, left and right panels) and different grid sizes (grid methods, right panels). The left panels show the results of computing for explicit schemes MKD SPH and ESPH. At $h = 0.025, h=0.01$, the time step $\tau = 0.001<t_{\rm stop}$ and the number of particles $N_{total} = 2 \times 600$ are used; while at $h=0.001$, the step $\tau = 0.0001$ and $N_{total} = 2 \times 6000$. The middle panels present the results of computing for semi-implicit schemes ISPH and SPH-IDIC with $CFL=0.1$; in this case, the number of particles was the same as for the left panels. The right panels illustrate the grid methods with the spatial steps $h=0.025, h=0.01$, and $h=0.001$; for the SIOS PPA method, the results at $h=0.0025$ are also shown. For both grid methods, $CFL=0.5$.

One can see that the numerical solutions obtained by MKD, ESPH and ISPH methods with the smoothing length increased from $h=0.001$ (the condition (\ref{eq:sparesSPH}) is satisfied) to $h=0.025$ ((\ref{eq:sparesSPH}) is violated) acquire a pronounced dissipation. The observed tendency to solution dissipation was described in  \cite{BateDust2014,LaibePrice2011}. Note that the maximum level of dissipation is obtained in the case of explicit schemes MKD and ESPH (at $h=0.025$ the calculated perturbation amplitude differs from the analytical one by more than 20\%). As follows from the central bottom panel in Fig.~\ref{fiq:schemes}, the semi-implicit ISPH scheme gives a smaller dissipation at the same smoothing length as compared to MKD and ESPH. Moreover, thanks to semi-implicit approximation of the drag force, the ISPH scheme has no restrictions on the time step (\ref{eq:timeresSPH}). MKD, ESPH and ISPH are the fully Lagrangian methods, which means that all forces are calculated without the introduction of a spatial grid. Evidently, of all these methods, ISPH is most suitable for disc simulation.

The SPH-IDIC method is a combination of Lagrangian and Euler approaches because the drag force is computed using the decomposition of particles into Euler volumes. One can see on the central upper panel of Fig.\ref{fiq:schemes} that these numerical solutions obtained by SPH-IDIC are free of dissipation, and at $h=0.025$ the wave amplitude is reproduced without visible error, in distinction to MKD, ESPH and ISPH. Thus, the computing of drag using the Euler grid makes it possible to obtain an acceptable accuracy of solutions even if conditions (\ref{eq:timeresSPH}) and (\ref{eq:sparesSPH}) are violated.      
As follows from the right upper panel of Fig.\ref{fiq:schemes}, similar SPH-IDIC results can be obtained by the grid method EMSI-PPA. The both methods are constructed so that in each computational cell the local momentum conservation law (\ref{eq:dragTransfer}) is fulfilled, which means that the momentum lost by gas due to dust drag is certainly equal to the momentum acquired by dust due to gas drag. In addition, a comparison of the computing results obtained by MKD, ESPH and ISPH schemes with SIOS PPA shows that non-fulfillment of (\ref{eq:dragTransfer}) due to operator splitting distorts not only the wave amplitude but also its propagation rate. 

\subsection{The role of back reaction from dust to gas in simulation the dynamics of a circumstellar disc}

To illustrate the performance of the methods under consideration, let us examine the results of modeling the dynamics of a circumstellar gas-dust disc within the FEoSaD model and taking into account a set of physical processes: viscous transport leading to the accretion of matter from gas onto the star, self-gravity of gas and dust components of the disc, gas heating and cooling, and dust growth. The numerical model of the gas component of the disc is described in detail in \cite{VorobyovBasu2015}. The numerical model of dust dynamics and growth is described in \cite{VorobyovEtAl2017}. Equations of gas and dust motion are integrated using the EMSI PPA method. 

\begin{figure}
\centering  
\includegraphics[scale=0.3]{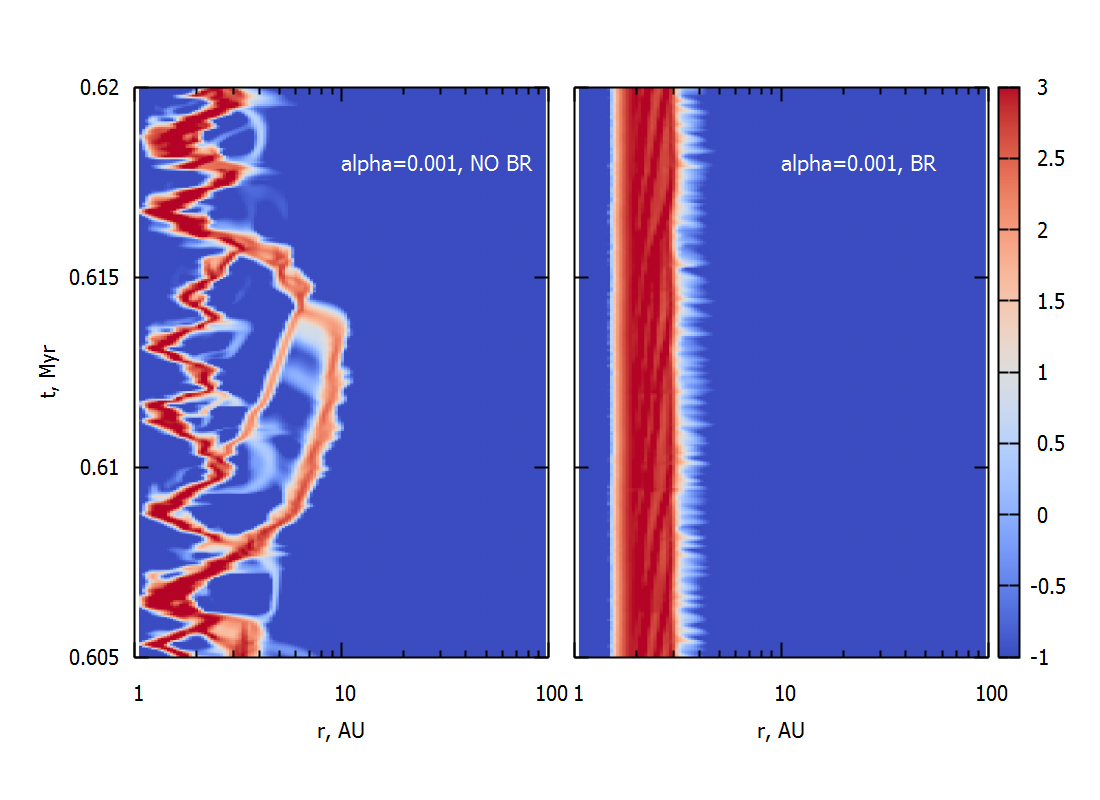}
\caption{The logarithm of the angle-averaged surface density of the grown dust in grams per square centimeter. Simulation with the FEoSaD \cite{VorobyovBasu2015} code using the EMSI-PPA method for computing of momentum transfer between gas and dust. Left panel - only one-way dust drag is taken into account, right panel - two-way momentum transfer is considered.}
\label{fig:RoleBR}
\end{figure}

Fig.~\ref{fig:RoleBR} displays the time dependence of the angle-averaged surface density of the growing dust in grams per square centimeter (a logarithm) for two models in which all physical and numerical parameters are identical, except the effect of dust drag on gas dynamics. The right panel corresponds to the model that takes into account the momentum transfer between gas and dust, while the left panel takes into account only the effect of gas on the dust dynamics. It is seen that in both models the grown dust is accumulated in the internal region of the disc. However, in the model where the momentum interchange is taken into account, dust concentrates in a dense ring, in the region where gas viscosity sharply changes (the so-called dead zone) and the pressure gradient sharply modifies. In the model that does not consider the back reaction of dust dynamics on the gas, the rings with a high concentration of dust rapidly move in the internal region of the disc. These effects will be elucidated in our future studies because the formation of regions with an increased concentration of the solid phase can initiate the formation of terrestrial planets and asteroidal belt.

\section{Conclusions}
\label{sec:resume}

The paper presents the algorithms for numerical simulation of the momentum transfer between solid phase and gas in multiphysical models of the circumstellar disc dynamics and planet formation. We found that SPH-IDIC and EMSI-PPA methods have the following properties:
\begin{itemize}
\item the universality (applicability to solids of different size and composition), 
\item the possibility to simulate the effect of dust of the gas dynamics, which occurs in the regions with a high concentration of dust, 
\item the compatibility with the scalable parallel algorithms for solving the spatially 3D  equations of gas dynamics and the Boltzmann equation. 
\end{itemize}

The reported algorithms are applicable at the following approximations of the physical model:  
\begin{itemize}
\item dust in the disc is assumed to be monodisperse, i.e. represented by one characteristic size,
\item the drag force linearly depends on the relative velocity between gas and solids (the Epstein regime, which is applicable to a wide range of sizes of solids in the disc), 
\item gas and solid phase transfer the momentum but not the energy, and 
\item the ratio of the velocity relaxation time $t_{\rm stop}$ to the revolution time of the disc (the Stokes number) does not exceed 10.
\end{itemize}

It was shown that in disc dynamics simulations taking into account the back reaction of dust drag on massive gas  may significantly affect the structure of gas and dust in the internal region of the disc, where terrestrial planets and asteroidal belt can be formed.

\ack
The work done by O. Stoyanovskaya, V. Akimkin, T. Glushko, Ya. Pavlyuchenkov and E. Vorobyov is supported by the Russian Science Foundation, Project 17-12-01168. The work done by N. Snytnikov was conducted within the budget project 0315-2016-0009 for ICMMG SB RAS, the work of V. Snytnikov was conducted within the budget project of Boreskov Institute of Catalysis SBRAS. Simulations were made at the Computing Center (ICMMG) of SB RAS, Lomonosov Computing Center of the Moscow State University, and Vienna Scientific Clusters VSC-2 and VSC-3. Development and testing of grid methods, as well as modification of FEoSaD code and modeling of disk dynamics were carried under the Russian Science Foundation, Project 17-12-01168.

\end{document}